\documentclass[a4paper,11pt]{article}
\usepackage[dvipdfmx]{graphicx}
\usepackage{pos}
\usepackage{graphicx}
\usepackage{threeparttable}

\newcommand\aap{A\&A}
\newcommand\apjl{ApJL}     

\newcommand\M{$M_{\rm{ZAMS}}$}

\title{Progenitor constraint using line ratios of the CNO elements  in supernova remnants}

\author*[a]{Hiroyuki Uchida}
\author[a]{Takuto Narita}

\affiliation[a]{Kyoto University,\\
  Kitashirakawa Oiwake-cho, Sakyo, Kyoto, 606-8502, Japan}

\emailAdd{uchida@cr.scphys.kyoto-u.ac.jp}

\abstract{
Unveiling the nature of progenitors is crucial for understanding the origin and the mechanism of core-collapse and thermonuclear supernovae (SNe).
While several methods have been developed to derive stellar properties so far, many questions remain poorly understood. 
In this paper we demonstrate an observational approach to constrain progenitors  of supernova remnants (SNRs) using abundances of carbon (C), nitrogen (N),  and oxygen (O) in  shock-heated circumstellar material (CSM). 
Our calculations with stellar evolution codes indicate that a total amount of these CNO elements will provide a more sensitive determination of the progenitor masses than the conventional method based on ejecta abundances.
If the CNO lines (particularly those of C and N) are detected and measured their abundance ratios accurately, they can provide relatively robust constraint on the progenitor mass (and in some cases the rotation velocity) of SNRs.
Since our method requires a better energy resolution and larger effective area in the soft X-ray band ($<1$~keV), XRISM launched on September 7, 2023 and next-generation microcalorimeter missions such as Athena, Lynx, LEM, and HUBS will bring a new insight into link between the progenitors and their remnants.

}

\FullConference{
Multifrequency Behaviour of High Energy Cosmic Sources XIV (MULTIF2023)\\
12-17 June 2023\\
Palermo, Italy\\}

\begin{document}
\maketitle

\section{Introduction} 
While  increasing diversity of supernova (SN) Types and subclasses has been reported so far, their progenitor origins and explosion mechanisms are still far from being well understood.
Recent theories and observations indicate that their circumstellar environments are the key to understanding such a variety of spectra and light curves.
It has been considered that several types of core-collapse (CC) SNe (e.g., Type IIn) can be related to a shock-heated circumstellar material (CSM)  due to a high mass-loss rate \citep[e.g.,][]{Smith_2014}; also, it has been recently realized that even in a regular Type II SN progenitors are commonly surrounded by the CSM that was evacuated just before the explosion \citep{Yaron_2017}.
Even in the case of Type Ia SNe, their circumstellar environments provide a clue to their origin since some of subclasses suggest massive CSM \citep[e.g.,][]{Dilday_2012} and more generally a low-density cavity might be formed by a mass-accreting white dwarf before the explosion \citep{Hachisu_1996, Badenes_2007}.
Therefore, observations of CSM around SNe are the key to understanding the explosion mechanisms of both CC and Type Ia SNe.

One of the most important topics in CC SNe is the progenitor mass distribution.
Several observations of extra-galactic SNe provide a rough estimation of  the mass range for explosion \citep[e.g.,][]{Smartt_2009, Diaz-Rodriguez_2018}.
Recent observations of supernova remnants (SNRs) indicate that the ejecta abundance ratio of iron (Fe) to silicon (Si) shows a negative dependence on the zero-age main-sequence masses (\M) \citep[][]{Sukhbold_2016, Katsuda_2018} and they categorized progenitor masses of SNRs into three cases: $15~M_{\odot}<M_{\rm{ZAMS}}$, $15~M_{\odot}<M_{\rm{ZAMS}}<22.5~M_{\odot}$, and $22.5~M_{\odot}<M_{\rm{ZAMS}}$.
While these studies are important for constraining \M \ that can explode as SNe, it is required a more precise measurement of the lower/upper limits of \M.
In this context, elemental abundances of shock-heated CSM in SNRs will provide rich information on  progenitor parameters: \M, an initial rotation velocity, convection, etc. \citep[e.g,][]{Fransson_2005, Maeder_2014, Chiba_2020}.
This is also true in the case of Type Ia SNe, because part of Type Ia SNRs such as Kepler's SNR and N103B indicate strong evidence for shock-CSM interactions \citep{Katsuda_2015, Kasuga_2021, Yamaguchi_2021, Guest_2022}.
The presence of the CSM may hint at the origin of the Type Ia SN explosion, for which two scenarios are under debate: the double-degenerate \citep[DD;][]{Iben_1984, Webbink_1984} or single-degenerate \citep[SD;][]{Whelan_1973}.
For instance, as previous studies indicate \citep[e.g.,][]{Reynolds_2007}, the presence of the nitrogen-rich CSM in Kepler's SNR support the SD scenario for the explosion.
If this is the case, it may be possible to investigate  the properties of a companion star, even though there is no promising candidate for a surviving star. 

Since stellar winds contain material processed  by the CNO cycle  in stellar interiors, light elements such as carbon (C), nitrogen (N),  and oxygen (O) most likely reflects  the progenitor (or a companion star's) properties.
It has been however difficult to detect the emission lines of especially C and N in shock-heated swept-up CSM with currently available  X-ray detectors.
This is mainly due to the lack of the energy resolution and effective area in the soft X-ray band, where the fluorescence lines of N~{\small VI}~Ly$\alpha$ (0.50~keV) and C~{\small V}~Ly$\alpha$ are expected.
While several observations report detections of C and N lines in SNRs with CCDs \citep[e.g.,][]{Miyata_2007}, they can be more clearly resolved with X-ray grating spectrometers, which have much better energy resolution \citep{Uchida_2019}.

We recently observed a magnetar-hosting SNR RCW~103 with the Reflection Grating Spectrometer (RGS)  on board XMM-Newton and measured the N abundance of the CSM in this remnant for the first time \citep{Narita_2023}.
By comparing the result with stellar evolution models, we successfully constrained \M \ and initial rotation velocity of the progenitor of RCW~103 to be 10--12~$M_{\odot}$ and $\lesssim 100$~km/s, respectively.
The method presented here will be useful  with upcoming X-ray microcalorimeter observations with XRISM \citep{Tashiro_2018} launched in 2023 and future missions such as Athena \citep{Barret_2018}, Lynx \citep{Gaskin_2019}, LEM \citep{Kraft_2022}, and HUBS \citep{Bregman_2023}.
We thus demonstrate  our calculation and method in this paper for near future high-energy-resolution X-ray observations of SNRs.

\begin{figure}[h!]
 \begin{center}
  \includegraphics[width=60mm]{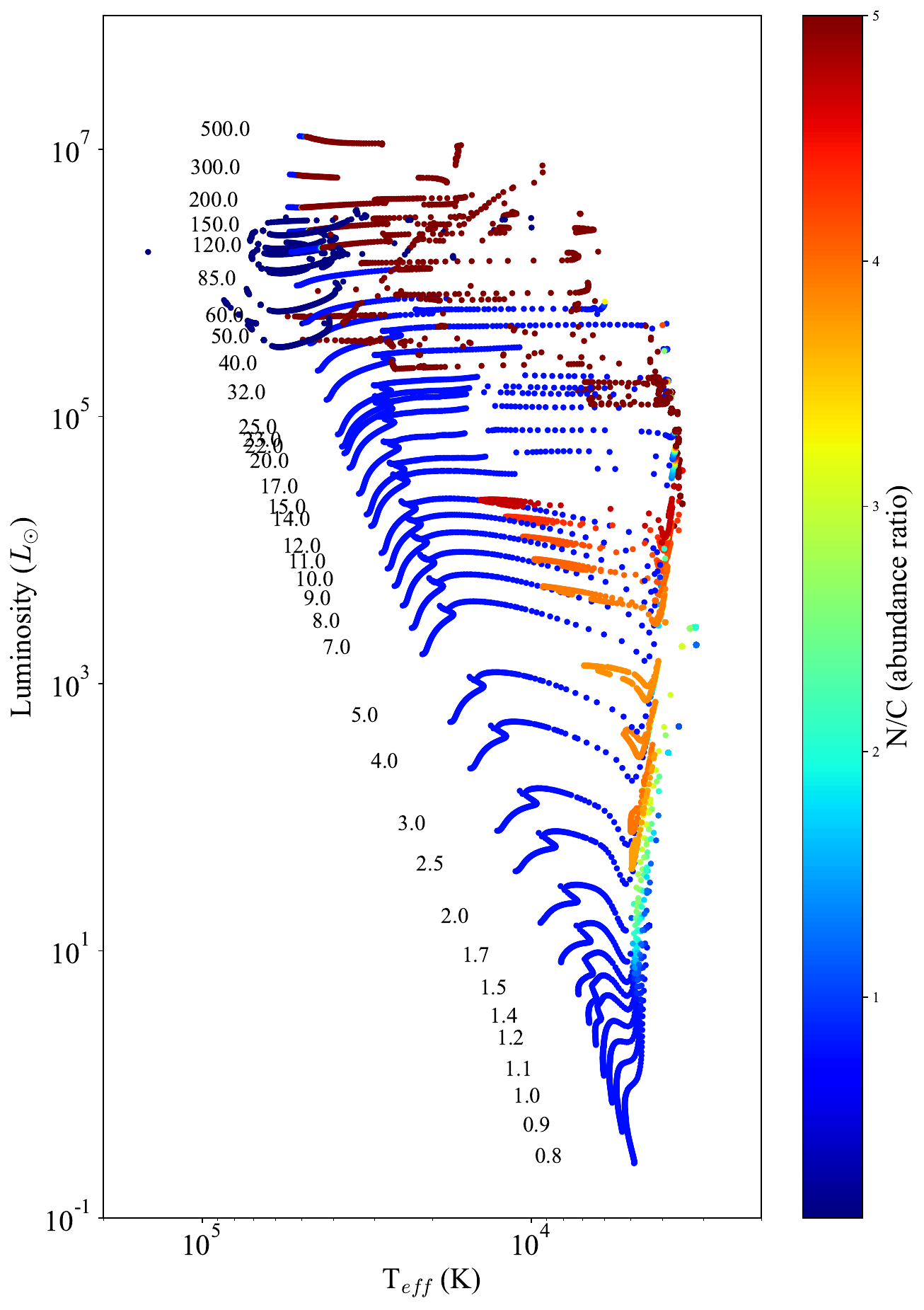}
  \includegraphics[width=60mm]{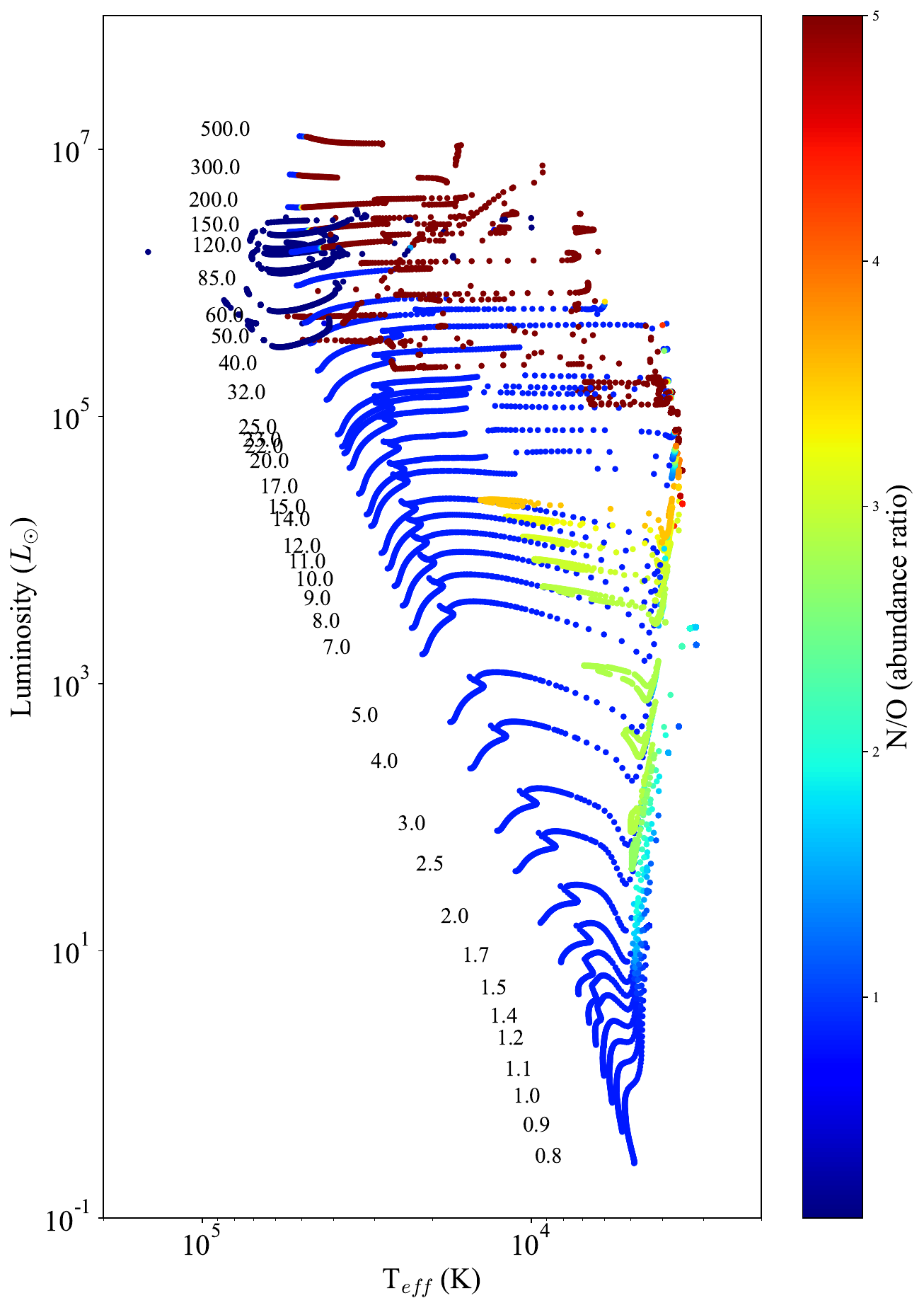}
  \includegraphics[width=60mm]{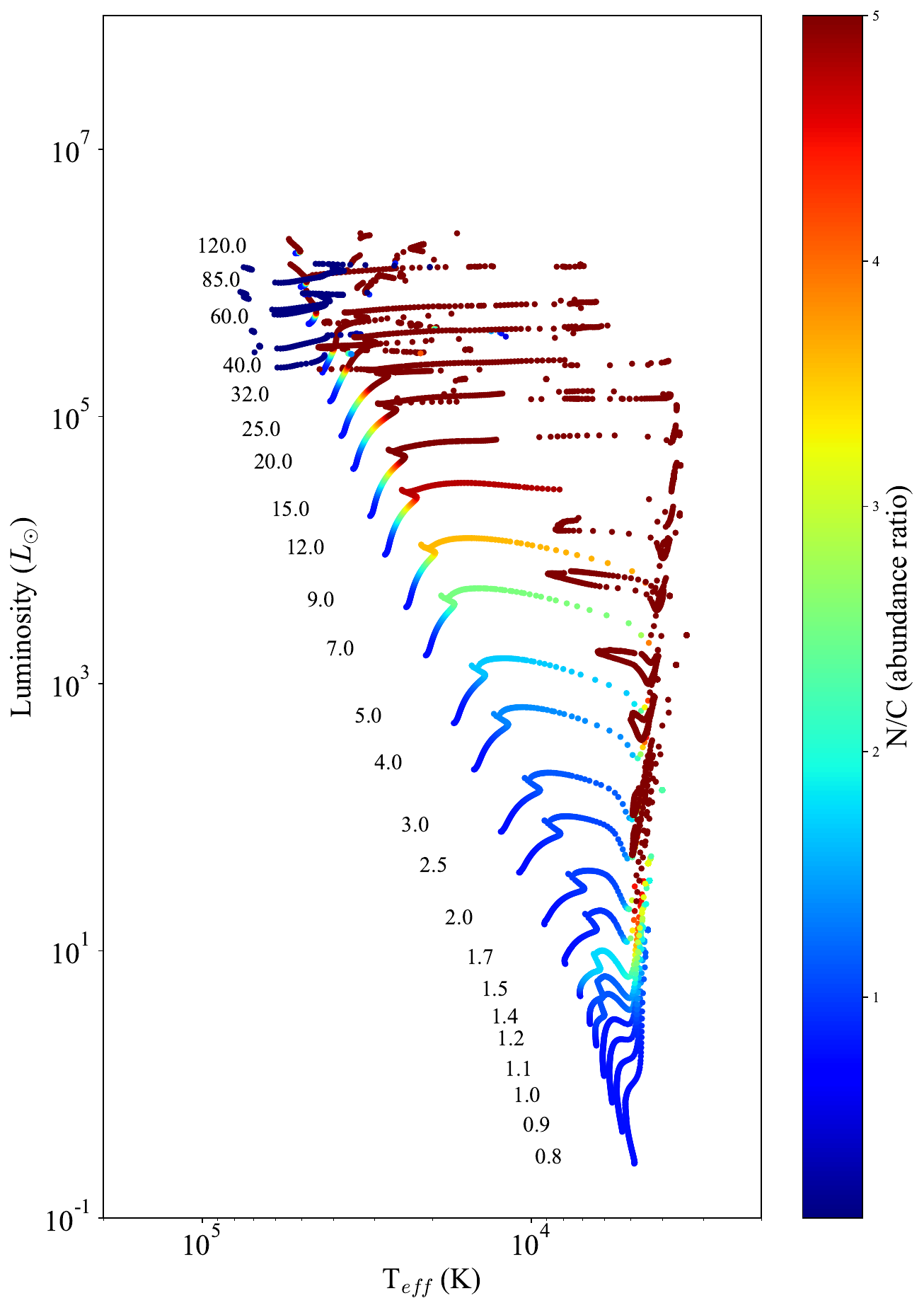}
  \includegraphics[width=60mm]{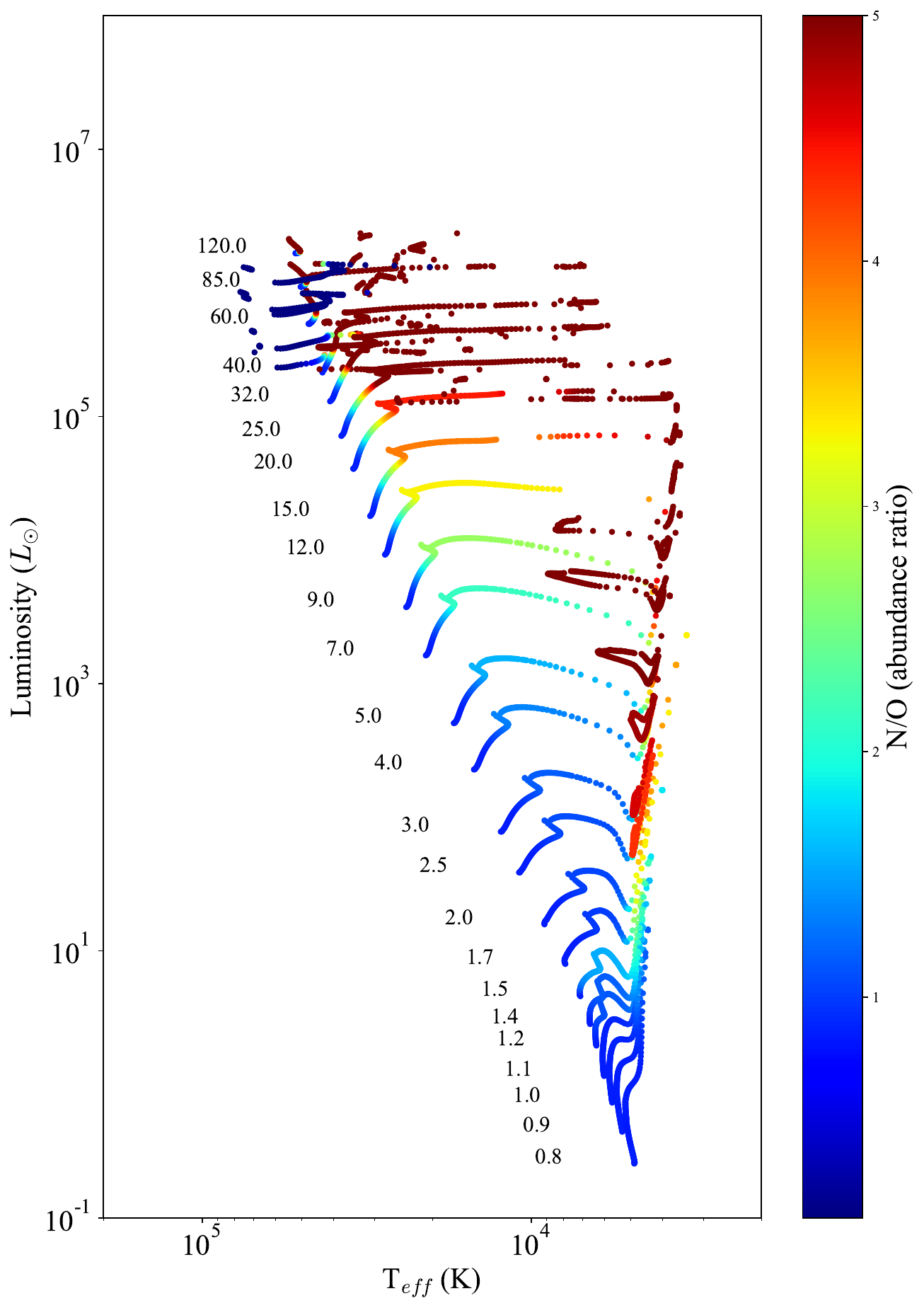}
 \end{center}
 \caption{Hertzsprung–Russell diagram and calculated abundance ratios for various stars with different \M, which are displayed at each start point of the main sequence. The color represents the abundance ratios of N/C (left) and N/O (right) that defined in equation~\ref{eq:ratio}. The upper and lower panels represent results for the initial rotation velocity $v_{init}/v_{\rm K}=0.0$ and 0.3, respectively.}
\end{figure} \label{fig:surface}

\section{Simulations and Calculations}
To calculate  expected abundance ratios for each progenitor mass, we used two stellar evolution codes: \textit{HOngo Stellar Hydrodynamics Investigator, HOSHI} \citep{Takahashi_2013, Takahashi_2014, Takahashi_2018, Yoshida_2019}  and \textit{Geneva}  \citep{Ekstrom_2012}.
Here we focus on total yields of light elements that produced by the CNO cycle in the stellar interiors during their evolution.
Since observationally it is useful to represent results in the form of so-called ``solar abundance ratio'', we define parameters as follows:
 \begin{equation}\label{eq:ratio}
{\rm N/O}\equiv\frac{(n_{\rm N}/n_{\rm H})/(n_{\rm N}/n_{\rm H})_{\odot}}{(n_{\rm O}/n_{\rm H})/(n_{\rm O}/n_{\rm H})_{\odot}}=\frac{(n_{\rm N}/n_{\rm O})}{(n_{\rm N}/n_{\rm O})_\odot},
\end{equation}
where $n_{\rm N}$, $n_{\rm O}$, and $n_{\rm H}$ represent the number density of each element, and $(n_{\rm O}/n_{\rm H})_{\odot}$ means the solar abundance ratio \citep[cf.][]{Wilms_2000}.
Note that we show only the case of solar metallicity for the following calculations.
This is because although total yields of the CNO elements highly depend on the stellar metallicity, their abundance ratios were almost unchanged and thus the following discussion is not significantly altered.

Figure~\ref{fig:surface} displays the Hertzsprung–Russell diagram and calculated abundance ratios of the CNO elements, which are produced inner layers and transferred to the stellar surface during each evolution stage with \M.
In every case, the abundance ratios are almost constant during the main sequence (MS) whereas they are increasing since N is accumulated at the post MS phase.
The result is reasonably consistent with the standard picture of the massive star evolution.
As also shown in the lower panels of Figure~\ref{fig:surface}, we changed the  initial rotation velocity $v_{init}/v_{\rm K}=0.0$ to 0.3, where $v_{\rm K}$ is the Kepler velocity at the surface.
As a result, in such fast-rotating star N is effectively evacuated and the resultant abundance ratios become larger.
Additionally it is also effective in some cases that the centrifugal force efficiently promotes the convection and nuclear reactions though this effect is relatively negligible.

\begin{figure}[h!]
 \begin{center}
  \includegraphics[width=100mm]{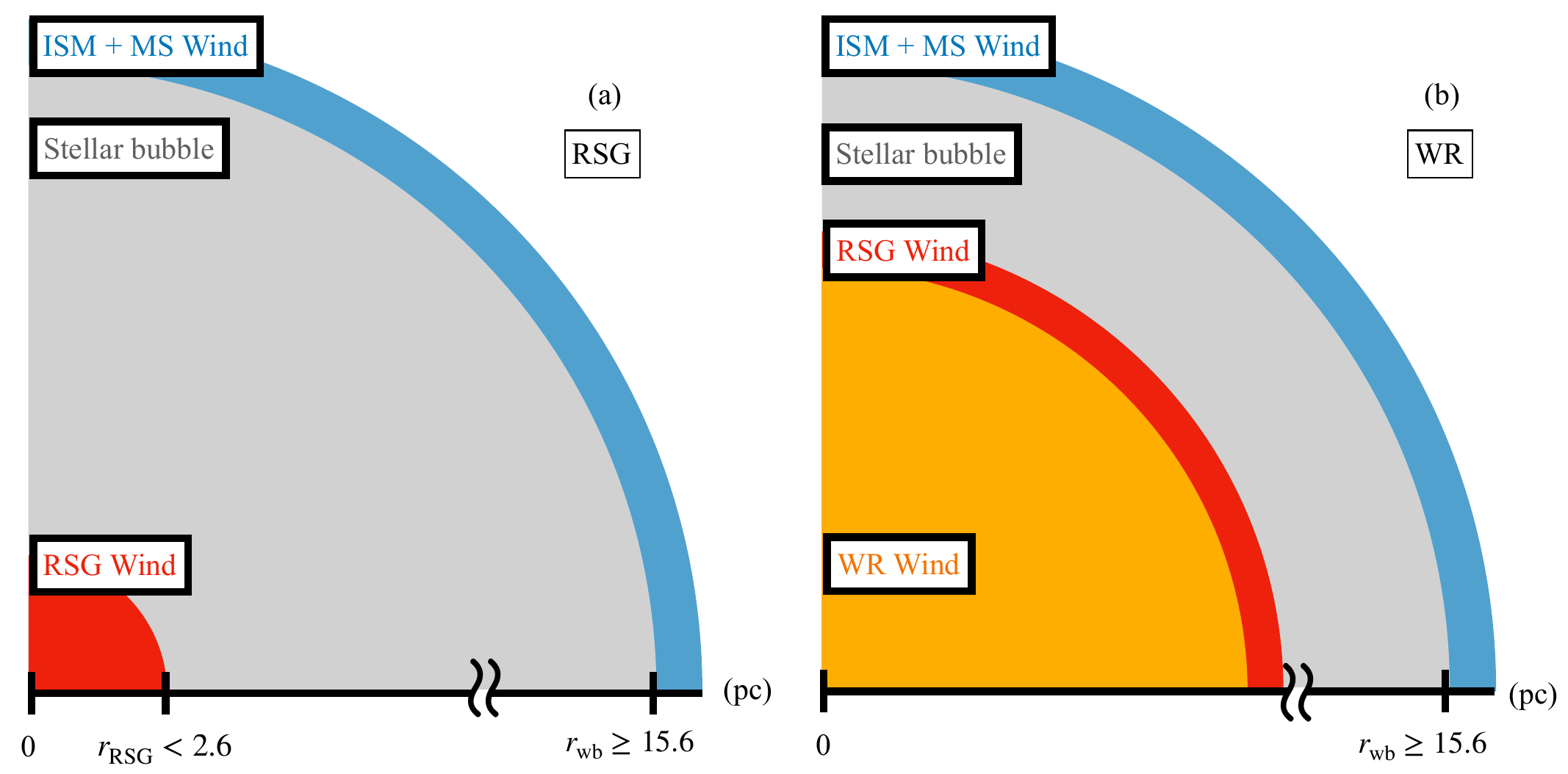}
 \end{center}
 \caption{Schematic view of an environment around a star at the end of its life. \textit{Left (a)}: Case of a low-mass star ending up as an RSG. Colors correspond to  the shell of ISM and MS winds (blue), stellar bubble (grey), and RSG winds (red). \textit{Right (b)}: Case of a high-mass star ending its life as a WR. The region dominated by a WR wind is represented as orange.}
\end{figure} \label{fig:structure}

The elements transferred to the stellar surface (Figure~\ref{fig:surface}) are mostly ejected by the stellar wind during each evolution phase.
Total amounts of the CNO elements in the CSM are therefore time-integrated values and determined by mass-loss history of MS, red supergiant (RSG), and in some cases Wolf-Rayet (WR) winds \citep{Weaver_1977, Chevalier_1983, Chevalier_2005}. 
If we consider the  abundance of shock-heated (swept-up) CSM for comparison with observations, the radial structure of their wind bubbles is also needed.
We thus assume a simple geometrical configuration as illustrated in Figure~\ref{fig:structure}.
For example, the total MS wind mass depends on the progenitor \M, which can be calculated with the HOSHI code.
The wind bubble size $r_{\rm{wb}}$ is typically calculated as
\begin{equation}\label{eq_rb_zams}
r_{\rm{wb}} =15.6\left(\frac{M_{\rm{w}}}{9.5\times10^{-1}~M_{\odot}}\right)^{1/3}
\left(\frac{v_{\rm{w}}}{10^{3}~\rm{km~s^{-1}}}\right)^{2/3}~\rm{pc},
\end{equation}
where $M_{\rm{w}}$ and $v_{\rm{w}}$ are the total mass and velocity of the MS wind, respectively (for more details, see \citep{Narita_2023}).
In the same manner, the size of the RSG wind  $r_{\rm{RSG}}$ is estimated to be 
\begin{equation}\label{eq_rR}
\begin{split}
r_{\rm{RSG}} \leq~&2.6\left(\frac{{\dot{M}}_{\rm{w}}}{5\times10^{-7}~M_{\odot}~\rm{yr^{-1}}}\right)^{1/2}\left(\frac{v_{\rm{w}}}{15~\rm{km~s^{-1}}}\right)^{1/2}~\rm{pc}.
\end{split}
\end{equation}
In the case of high-mass stars, a WR wind expands into the RSG-wind materials with a velocity of 100--200~km~s$^{-1}$ \citep{Chevalier_2005} during a WR phase duration of $\sim10^{3\textrm{--}4}$ year for \M~$\lesssim60~M_{\odot}$.
Taking into account these properties, we illustrate a typical schematic view of a pre-SN environment in Figure~\ref{fig:structure}.
Since abundances of the CSM depend on where the SNR forward shock reached in the bubble, we consider three cases with different SNR radii $R_{\rm SNR}$, 1~pc, 5~pc, and 10~pc for the following calculations.

\begin{figure}[h!]
 \begin{center}
  \includegraphics[width=68mm]{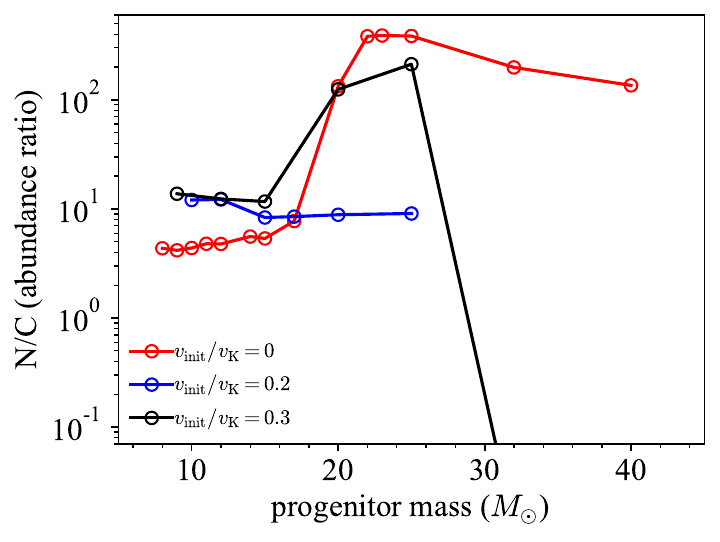}
  \includegraphics[width=68mm]{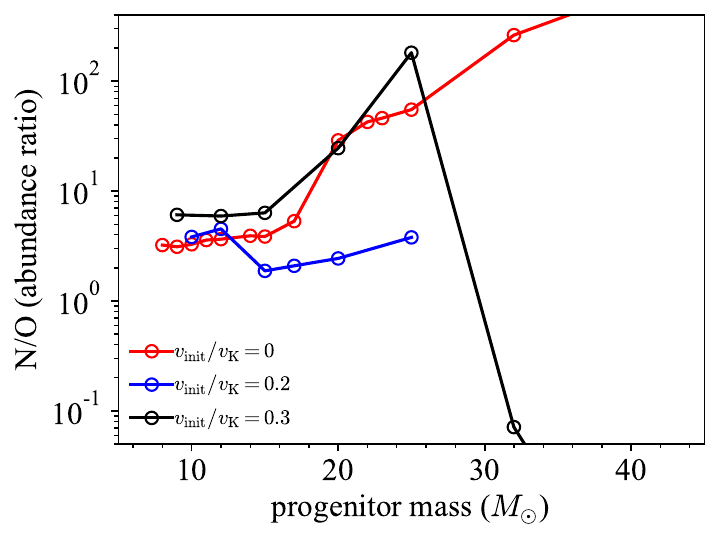}
  \includegraphics[width=68mm]{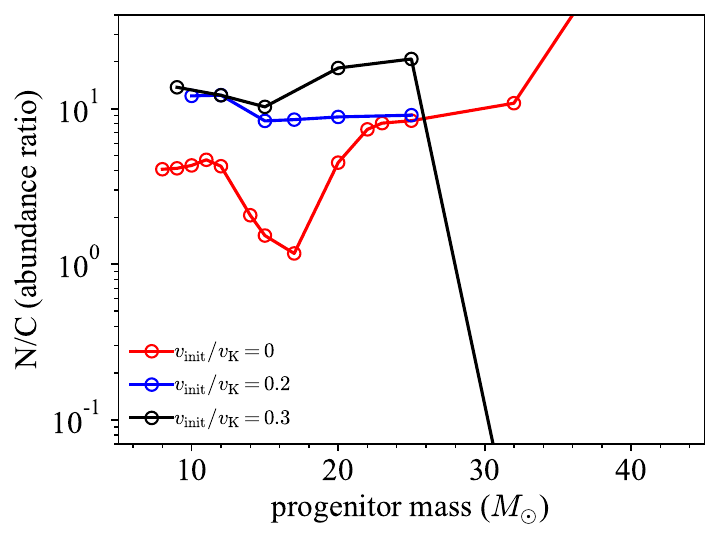}
  \includegraphics[width=68mm]{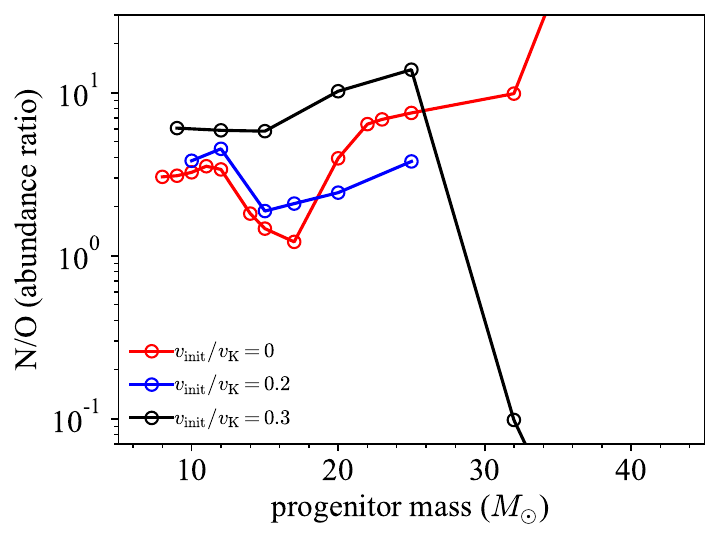}
  \includegraphics[width=68mm]{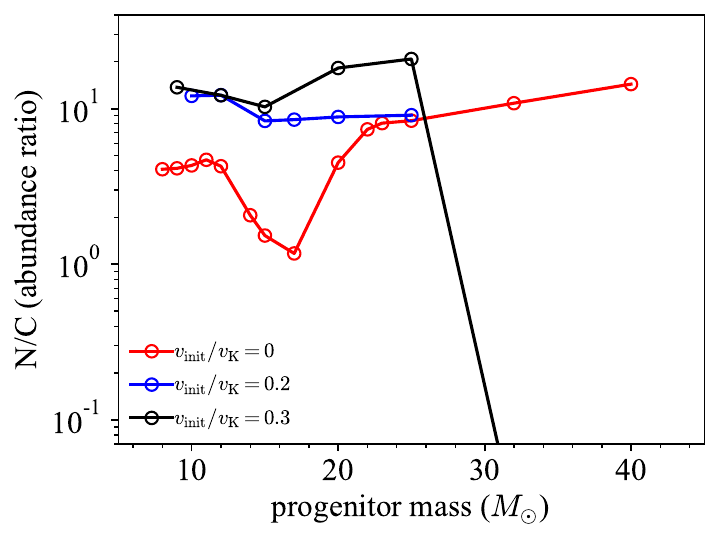}
  \includegraphics[width=68mm]{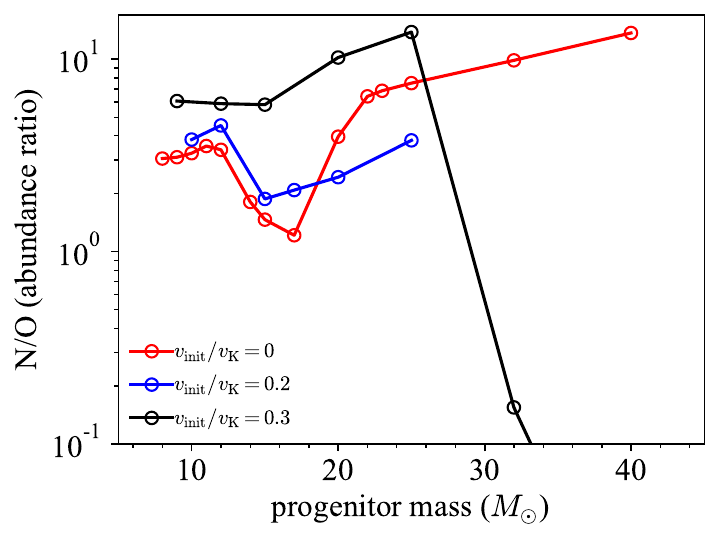}
 \end{center}
 \caption{Abundances ratios (see equation~\ref{eq:ratio}) of N/C (the left panels) and N/O (the right panels) expected  in the swept-up CSM with different \M \ calculated with HOSHI and Geneva codes. 
 The red, blue, and black circles correspond to  cases of $v_{init}/v_{\rm K}=0.0$, 0.2, and 0.3, respectively.
 The top, middle, and bottom panels indicate that the shell size of the remnant is assumed to be 1~pc, 5~pc, and 10~pc in radius, respectively.}\label{fig:abun}
\end{figure} 

\section{Results}
\begin{figure}[h!]
 \begin{center}
  \includegraphics[width=74mm]{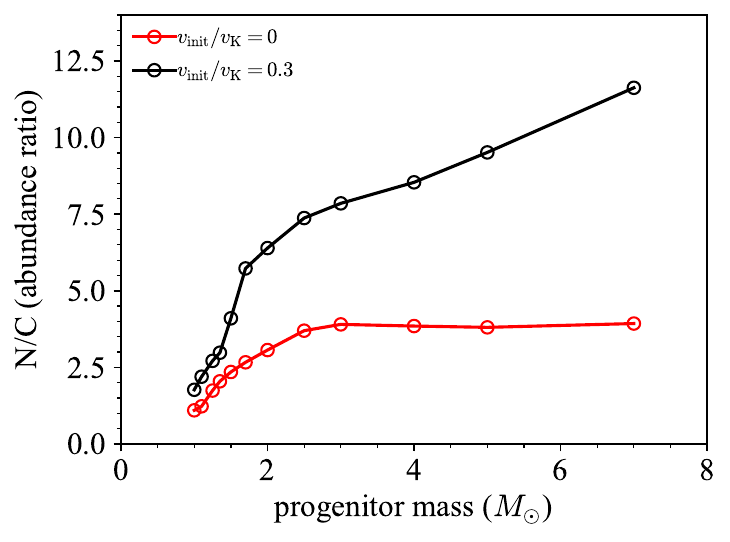}
  \includegraphics[width=70mm]{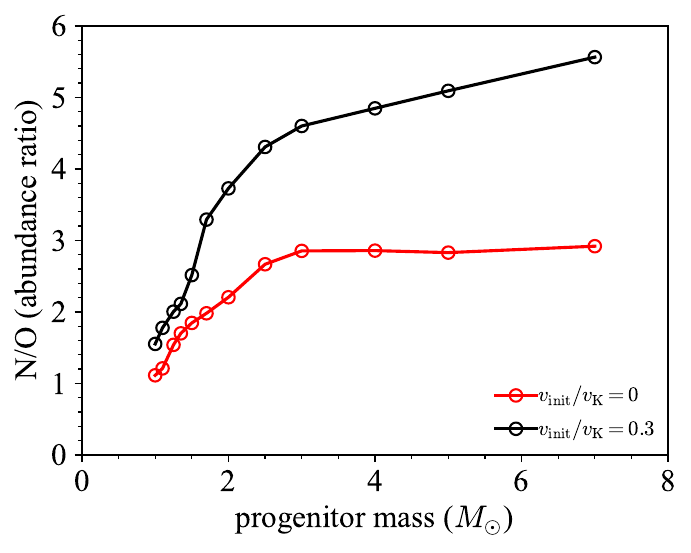}
 \end{center}
 \caption{Same as Figure~\ref{fig:abun}, but for lower-mass stars.
 The red and black circles correspond to  cases of $v_{init}/v_{\rm K}=0.0$ and 0.3, respectively.
 }\label{fig:abun_low}
\end{figure}

Figure~\ref{fig:abun} shows calculated abundance ratios of N/C and N/O assuming the three cases with different radii explained above.
We also calculated those for lower mass stars that do not explode as CC SNe as displayed in Figure~\ref{fig:abun_low}.
Note that the same plots of ``mass ratios'' as the x-axis are also given in Figures~\ref{fig:mass} and \ref{fig:mass_low} in Appendix for convenience.
MS stars emit a hydrogen(H)-rich wind whereas post-MS stars emit a stellar wind rich in helium (He) and N due to the convection, which carries the N-rich material from the the H-burning layer to the stellar surface \citep[e.g,][]{Owocki_2004, Przybilla_2010}.
In general, the total amount of  ejected N positively correlated to \M \ since the mass-loss rate has a positive relationship with the stellar mass \citep[e.g.,][]{Mauron_2011}.
Therefore, an increasing trend of N/C and N/O can be clearly seen for lower-mass stars that end their lives as red giant (RG) stars ($\lesssim3M_\odot$; Figure~\ref{fig:abun_low}).
A similar trend is also roughly seen in the case of $\gtrsim20M_\odot$, $v_{init}/v_{\rm K}=0.0$ (Figure~\ref{fig:abun}).
On the other hand, if the mass-loss rate is extremely high in a higher mass star, the N-rich wind is blown out in an earlier stage of evolution because of a stronger radiative force, so that N-rich CSM may be relatively far away from the progenitor; the inner C- and O-rich layer is emitted afterward.
In this case, the resultant abundance ratios of N/C and N/O become lower.
The  entire trend therefore appears  more complicated.

Another effect that changes the abundance ratios is especially considerable for lower-mass stars.
Stars with \M$\simeq4$--8$M_{\odot}$ end up as an asymptotic giant (AGB) branch \citep[e.g.,][]{Siess_2010}, which results in a rapid increase of N at the early stage of the post MS phase: N-rich materials in the H-burning layer are carried to near the stellar surface by the so-called dredge-up \citep{Karakas_2014}.
As a result, most of the stellar wind material is ejected from the single N-rich layer regardless of \M, which leads to the  almost constant trend in N/O.
In the case of stars with \M$=10$--12$M_{\odot}$, N-rich materials are carried to the surface  by convection; in some cases the He-burning occurs before the RSG phase, which prevents the star's expansion and hence an inefficient transfer of N.

The rotation velocity is also a considerable parameter; a faster rotation results in a more rapid increase of N during the stellar evolution.
This is because the  centrifugal force enhances the mass-loss rate at the stage of OB stars \citep{Owocki_2004}.
Note that additional N production by the CNO cycle is also enhanced in the post MS phase \citep{Heger_2000}, which is however less effective.
The effect of the stellar rotation can be typically seen in the cases of \M$\lesssim20M_\odot$ in Figures~\ref{fig:abun} and \ref{fig:abun_low}.

\section{Discussion}

\subsection{Core-collapse SNRs}
If the CNO lines are detected in an SNR and their abundances are measured,  the plots shown in Figure~\ref{fig:abun} may be a good tool for estimating the progenitor properties of CC SNe more accurately than before, as is demonstrated by our previous work \citep{Narita_2023}.
The most general method for this purpose is comparing an abundance ratio of (heavy) elements in ejecta with theoretically expected yield of explosive nucleosynthesis products.
Recent study \citep{Katsuda_2018} however points out that this conventional method suffers a large uncertainty except for the Fe/Si ratio, which is relatively sensitive to progenitor core masses.
Since all such methods using the ejecta abundance  highly depend on supernova explosion theory, which is currently  under progress, the mass estimation accuracy is less precise and provide a less stringent constraint on other parameters such as the rotation velocity.
In contrast, the drawback of our method is the  difficulty in detection of the CNO lines with currently available detectors, which should be overcome with future mission with a higher energy resolution and a larger effective area in the soft X-ray band ($<0.7$~keV).

\begin{figure}[h!]
 \begin{center}
  \includegraphics[width=75mm]{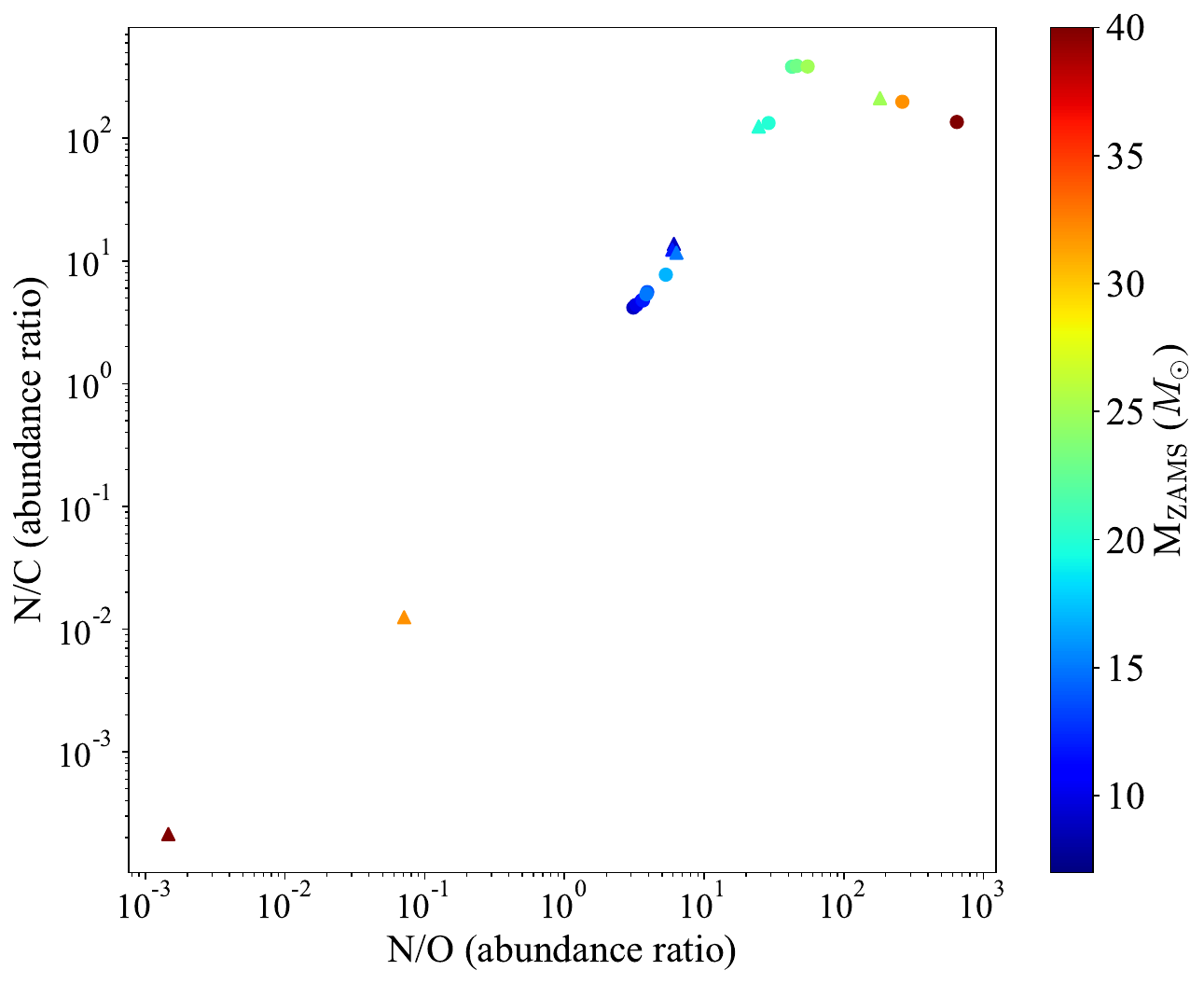}
  \includegraphics[width=75mm]{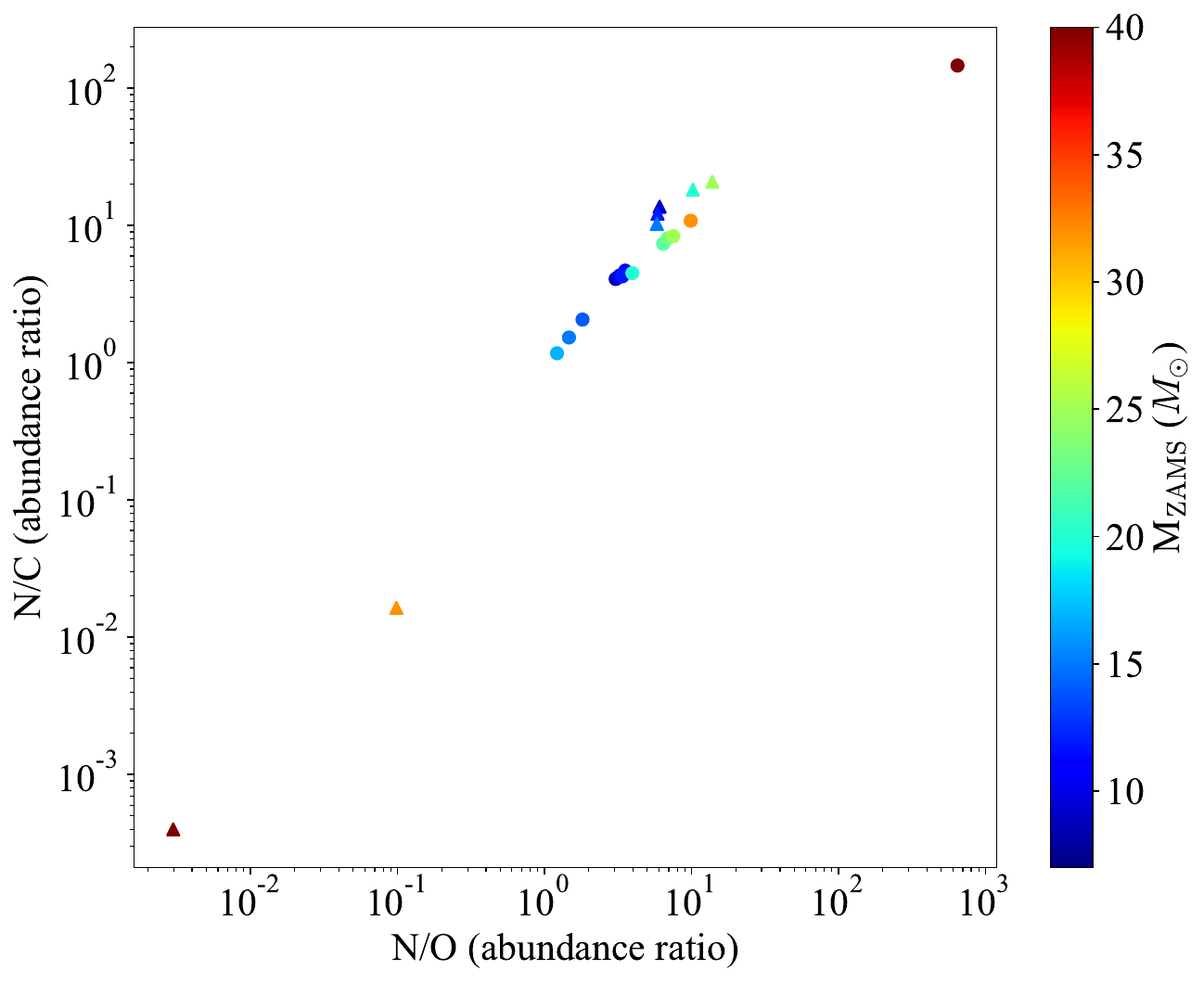}
  \includegraphics[width=75mm]{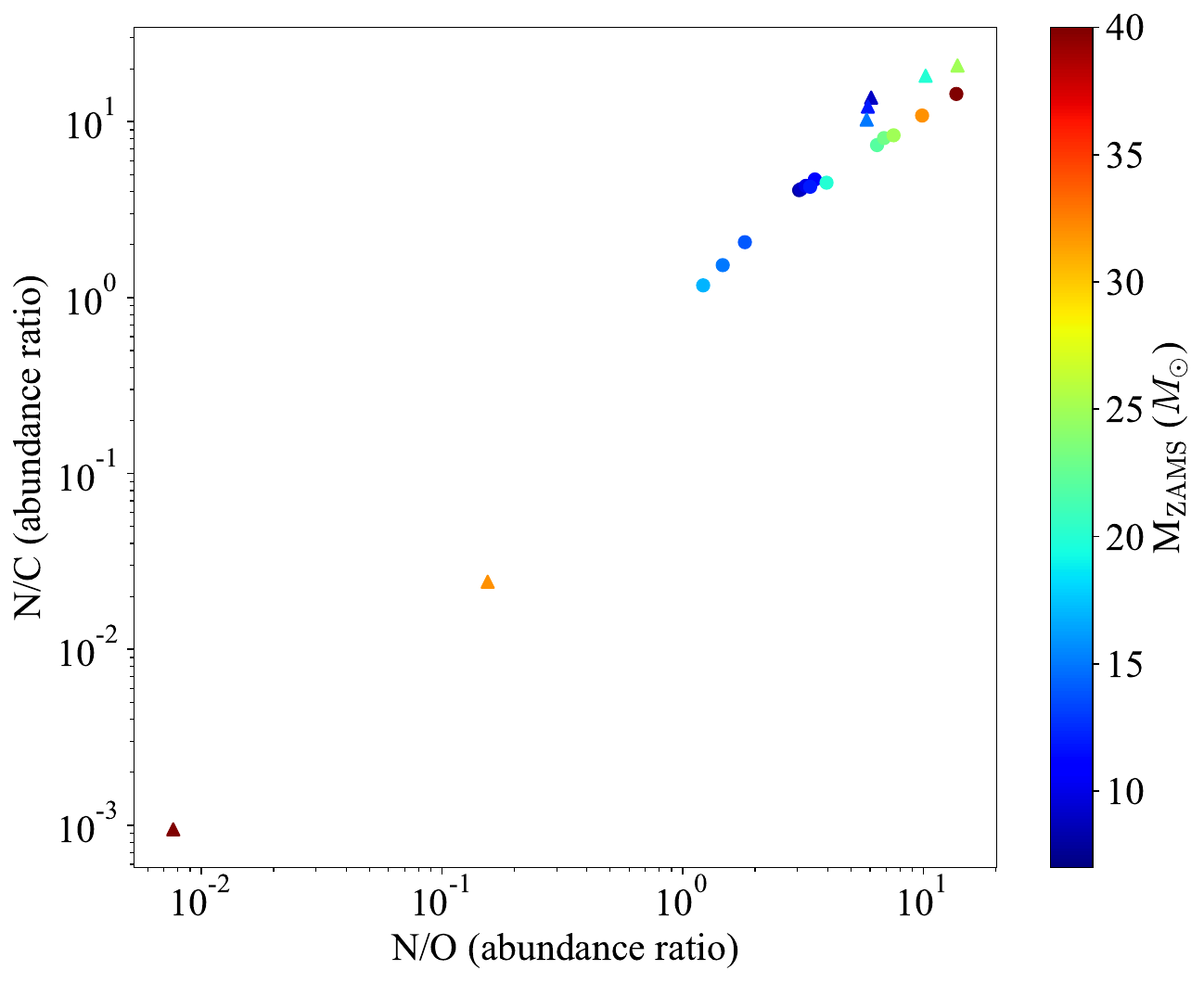}
 \end{center}
 \caption{N/C versus N/O abundance ratios. The colored points indicate ratios obtained in our calculations. The circles and triangles represent cases of $v_{init}/v_{\rm K}=0.0$ and 0.3, respectively.
 The top left, top right, and bottom panels indicate that the shell size of the remnant is assumed to be 1~pc, 5~pc, and 10~pc in radius, respectively.
 }\label{fig:ncno}
\end{figure}

Figure~\ref{fig:ncno} displays a N/C--N/O diagram based on our results shown in Figure~\ref{fig:abun}.
A clear correlation is found between the CNO abundances and the progenitor \M.
It seems therefore promising to constrain \M \ of SNRs by measuring both of N/C and N/O from a CSM spectrum, while C~{\small V}~Ly$\alpha$ (0.37~keV) is hard to detect in SNRs except for several examples \citep{Uchida_2019}.
If a systematic spectroscopy of SNRs is performed  with a microcalorimeter such as \textit{Resolve} onboard XRISM, its scattering plot will provide a clue to the progenitor mass distribution, which is observationally less understood \citep[e.g.,][]{Williams_2018}.
A caveat of using this method is that a uniform ambient density is tacitly assumed for calculating the size of the wind bubble; if an SNR is strongly interacting with dense clouds, it makes a larger uncertainty in the abundance estimates.
Even if it were, this method will be still useful to estimate progenitor properties.
 
 If the progenitor is in a binary system before its explosion, expected CNO abundance can be significantly changed and therefore our method cannot be used directly.
 In this case we can first safely assume that the star experience a much stronger mass loss than that of a single star.
 Since inner H/He-burning layers tends to be exposed more effectively \citep[e.g.,][]{Smith_2014}, the progenitor star will end up as a WR star rather than a RSG star  by enhanced mass stripping \citep{Yoon_2010, Yoon_2017}.
 We predict that such mass stripping results in a relatively low N/O in the CSM because O-rich material is blown off from the exposed surface by a WR wind.
 While it is generally  hard to make a qualitative constraint and required an elaborate stellar wind simulation, one can roughly estimate the total amount of stripped envelope from the observed N/O.
By comparing the result with binary evolution simulations, the progenitor (and companion star's) mass is roughly determined.
A more detailed discussion will be given in our next paper \citep{Narita_2024}, in which we found evidence for the binary remnant in a Galactic SNR.

\subsection{Type Ia SNRs}
Several observations of Type Ia SNRs suggest the presence of shock-heated dense gas, which is thought to be the CSM or a wind bubble  created by an accreting progenitor white dwarf \citep[e.g.,][]{Katsuda_2015, Tanaka_2021}.
As indicated in Figure~\ref{fig:abun_low}, the result shows a monotonous increase in the amount of N with increasing \M.
These ratios are thus also  potentially useful for investigating the progenitor parameters of a companion star  in a binary system, and are particularly important in the context of the origin of Type Ia SNe.
Notably all the results significantly exceed the solar abundances; these ratios thus probe whether a forward-shock heated plasma is the CSM origin or interstellar matter (ISM) origin, which can be a good indicator for distinguishing between SD and DD progenitor scenarios.
We note that in our calculation  total amount of the ejected CNO elements is integrated from the MS to the final stage (e.g., RG or AGB) of stellar evolution.
Although a mass-accreting wind expected for the SD system \citep{Hachisu_1996} makes the situation more difficult than the case of CC SNe, it is possible for our method to change the integration time in the same manner shown in \citep{Narita_2023}, which may enable us to estimate plausible channels of the binary system such as WD/MS or WD/RG in a future analysis.


\begin{table*}[t]
\caption{SNRs, in which abundances of the CNO elements were measured by previous studies so far.}
\begin{center}
\begin{tabular}{lcccccc}
\hline
\hline
 Name   & Age  & C & N & O & N/O & ref. \\
\cline{3-5}
& (yr) & \multicolumn{3}{c}{(solar abundance)} & &  \\
\hline	   						    		      			  						
\textit{Galactic SNRs:} &&&&&\\
 ~~~Kepler's SNR  & $4\times10^2$ & (=O) & $3.3^{+0.2}_{-0.3}$ & 1 (fixed)  & $3.3^{+0.2}_{-0.3}$ & \citep{Katsuda_2015} \\ 
 ~~~RCW~103  & $2\times10^3$ & (=O)&  $1.5\pm0.1$ & $0.42^{+0.03}_{-0.01}$ & $3.8\pm0.1$ & \citep{Narita_2023} \\ 
 ~~~G296.1$-$0.5 & $2\times10^4$ & (=O)&  $0.63\pm0.07$ & $0.17\pm0.02$ & $\sim3.7$ & \citep{Tanaka_2022}\\ 
~~~Cygnus Loop  & (1--2)$\times10^4$ & $0.20\pm0.02$ &  $0.26^{+0.18}_{-0.02}$ & $0.18^{+0.02}_{-0.01}$ & $\sim1.4$ & \citep{Uchida_2019} \\ 
\hline
\textit{SMC/LMC SNRs:} &&&&&\\
 ~~~SN~1987A  & $4\times10$ & 0.12 & $1.05^{+0.16}_{-0.15}$ & $0.13\pm0.02$ & $\sim8.1$ & \citep{Lundqvist_1996, Sun_2021} \\ 
 ~~~N132D  & $2.5\times10^3$ & $0.26^{+0.02}_{-0.01}$ &  $0.172^{+0.009}_{-0.001}$ & $0.34^{+0.01}_{-0.02}$ & $\sim0.5$ & \citep{Suzuki_2020} \\ 
 ~~~DEM~L71  & $7\times10^3$ & $2.3\pm0.8$ &  $0.35\pm0.21$ & 1 (fixed) & $0.35\pm0.21$ & \citep{Heyden_2003} \\ 
 ~~~N23  & $8\times10^3$ & $0.18\pm0.03$ & $0.07\pm0.02$ & $0.26\pm0.01$ & $\sim0.3$ & \citep{Broersen_2011} \\ 
~~~J0453.6$-$6829  & (1.2--1.5)$\times10^4$ & (=O)&  $1.5\pm0.1$ & $0.42^{+0.03}_{-0.01}$ & $\sim0.7$ & \citep{Koshiba_2022} \\ 
 \hline
\end{tabular}\label{tab:snrs}
\end{center}
\end{table*}

\subsection{Current Observations and Future Prospects}
Due to the lack of the energy resolution and the effective area of current X-ray detectors, samples that can be detected the CNO (especially C and N) lines are limited to those of less absorbed targets such as nearby or high galactic latitude SNRs.
Table~\ref{tab:snrs} shows a (not exhaustive) list of  SNRs, in which the CNO abundances were measured with grating spectrometers so far.
We found that all the Galactic SNRs have a high abundance ratio of N/O: at least two samples, RCW~103 and Kepler's SNR, are previously reported of a possible CSM interaction \citep{Frank_2015, Kasuga_2021}.  
Interestingly, SNRs in the large/small Magellanic clouds (LMC/SMC) listed in Table~\ref{tab:snrs} have significantly lower N/O than the solar abundance ratio except for SN~1987A.
The extremely high N/O found in SN~1987A is not surprising because  there are many pieces of evidence for a shock-heated stellar wind material blown from the progenitor's RSG phase and the subsequent blue supergiant phases \citep{Lundqvist_1991, Burrows_1995}.
In such case, the N-rich wind blown just before the SN explosion is dominant in the swept-up shock-heated CSM, which gives a higher N/O.
On the other hand, much lower N/O found in the other LMC/SMC SNRs would be notable.
It cannot be explained by our model  displayed in Figure~\ref{fig:abun}.
Although a more massive star case (\M \ of $\gtrsim 30~M_{\odot}$) is acceptable,  it is unlikely in terms of a mass range that can explode as an SN \citep[$\lesssim 20~M_{\odot}$;][]{Sukhbold_2016}.
While the reason of the  observed low N/O is not obvious,  two possibilities are considered.
One is the observational difficulty in distinguishing between the CSM and ejecta components in their X-ray spectra.
Since these measurements were performed with RGS, it is technically hard to select only CSM regions that are expected to be in outermost shells.
If the spectra is contaminated by  ejecta emission and cannot be identified, it may result in an underestimate of the O abundance.
Another possibility is an effect of binary evolution.
Even in the case of a lower  \M \ progenitor ($\lesssim 15~M_{\odot}$), the mass-loss rate is enhanced in a binary system and a lager contribution of O-rich winds from He-burning products significantly reduces N/O compared with the single star case \citep[see also,][]{Narita_2024}.
This possibility is suggestively consistent with  present binary population studies, in which high-mass X-ray binaries are more numerous in low-metallicity galaxies \citep{Dray_2006, Douna_2015}.
It is required future follow-up investigations with larger sample of SNRs using a microcalorimeter onboard XRISM \citep{Tashiro_2018} and forthcoming missions such as Athena \citep{Barret_2018}, Lynx \citep{Gaskin_2019}, LEM \citep{Kraft_2022}, and HUBS \citep{Bregman_2023}.

\section{Summary}
Since stellar winds contain elements produced during stellar evolution, their abundances, especially those of C, N, and O reflect the progenitor properties.
In SNRs,  wind-blown material is observed as surrounding shock-heated CSM, and therefore the detection of the CNO lines is crucial for obtaining direct information on their progenitors.
Under this view, we calculated total amount of the CNO elements in the CSM using currently available stellar evolution codes.
The result indicates that the abundance ratios of N/O and N/C are sensitive to the progenitor \M \ and rotation velocities: these parameters can be a good probe to constrain the progenitor properties.
A pilot study has been done by our previous observation \citep{Narita_2023} and another study will be published soon \citep{Narita_2024}.
While SNRs in which previous observations detected the C and N lines are still limited, we found that N is more overabundant  than O (${\rm N/O}\sim3$) in most of galactic remnants whereas N/O is less than the solar abundance in those in the SMC and LMC except for SN~1987A.
The reason for this discrepancy is still unclear though it may be attributed to the different metallicity environments.
A more comprehensive systematic analysis will be required for many galactic/extra-galactic SNRs with  the state-of-the-art instruments like \textit{Resolve} on board XRISM.

\appendix
\section{Appendix}
\begin{figure}[h!]
 \begin{center}
  \includegraphics[width=68mm]{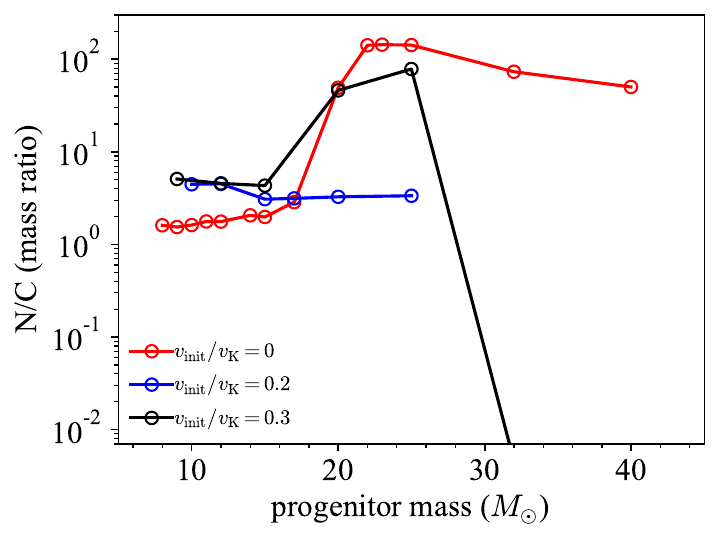}
  \includegraphics[width=68mm]{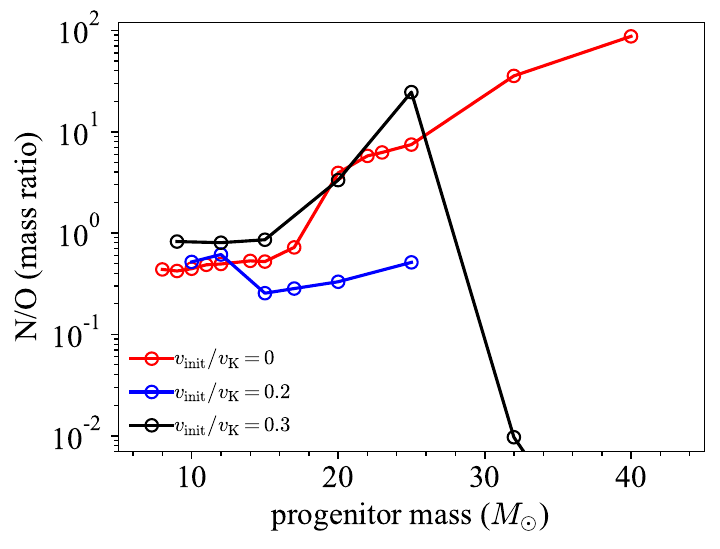}
  \includegraphics[width=68mm]{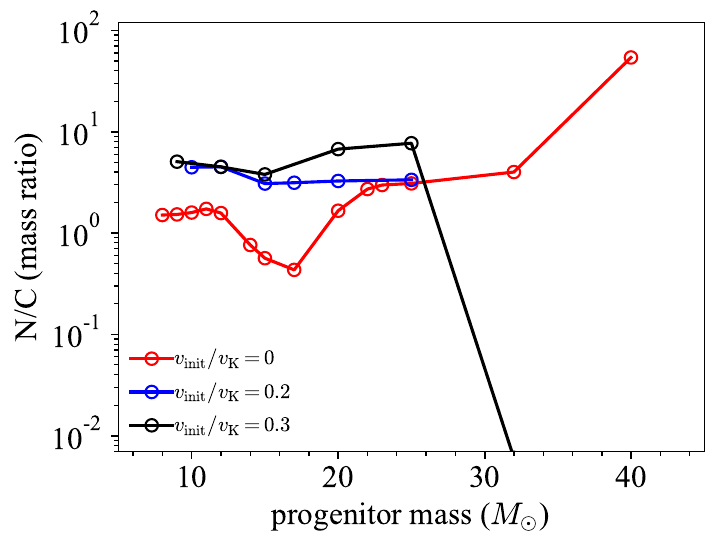}
  \includegraphics[width=68mm]{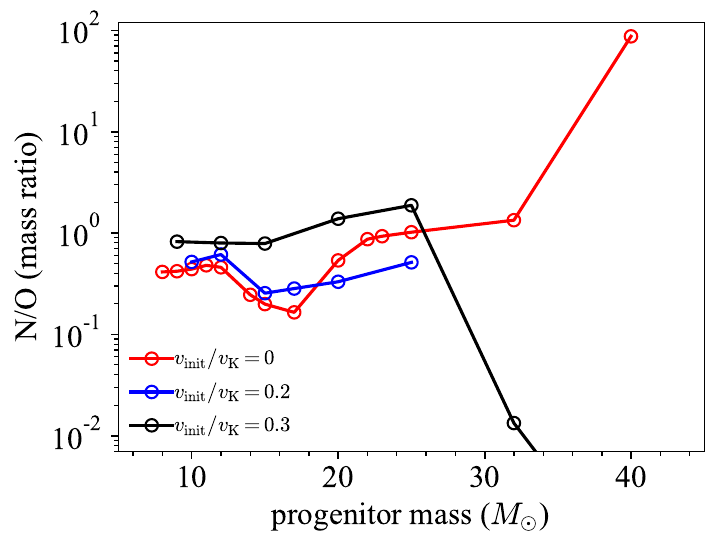}
  \includegraphics[width=68mm]{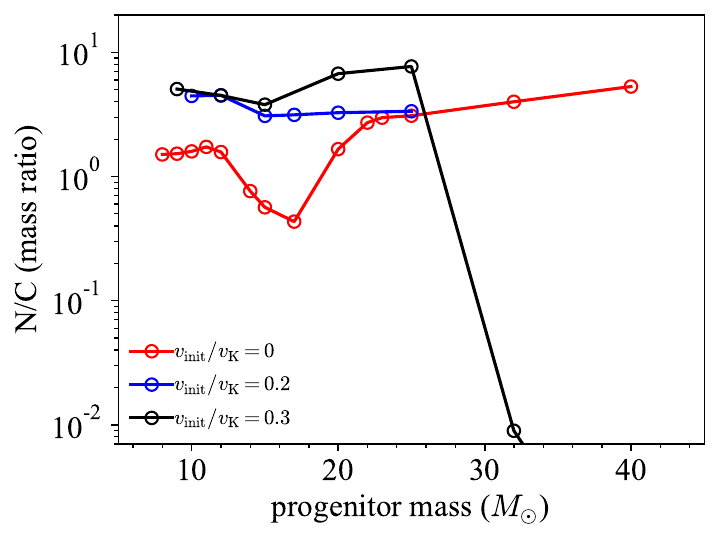}
  \includegraphics[width=68mm]{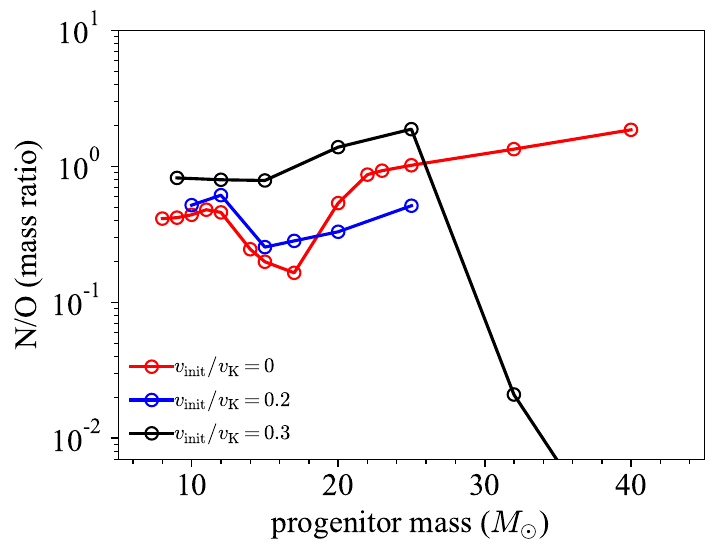}
 \end{center}
 \caption{Same as Figure~\ref{fig:abun}, but the x-axis is mass ratios.}\label{fig:mass}
\end{figure}

\begin{figure}[h!]
 \begin{center}
  \includegraphics[width=70mm]{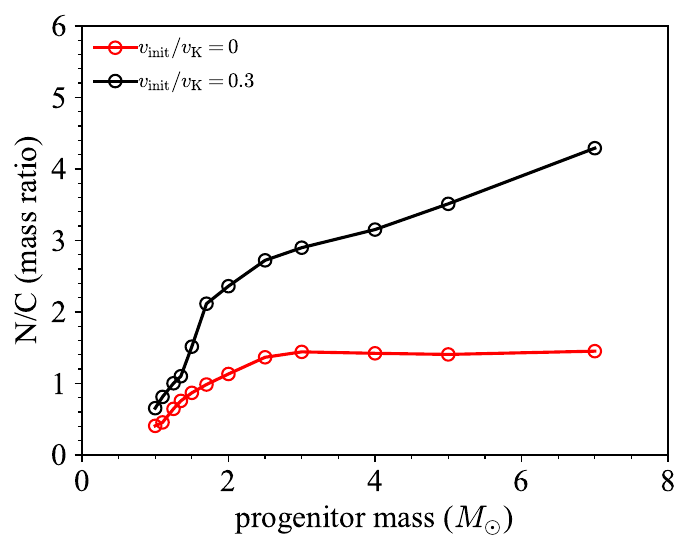}
  \includegraphics[width=70mm]{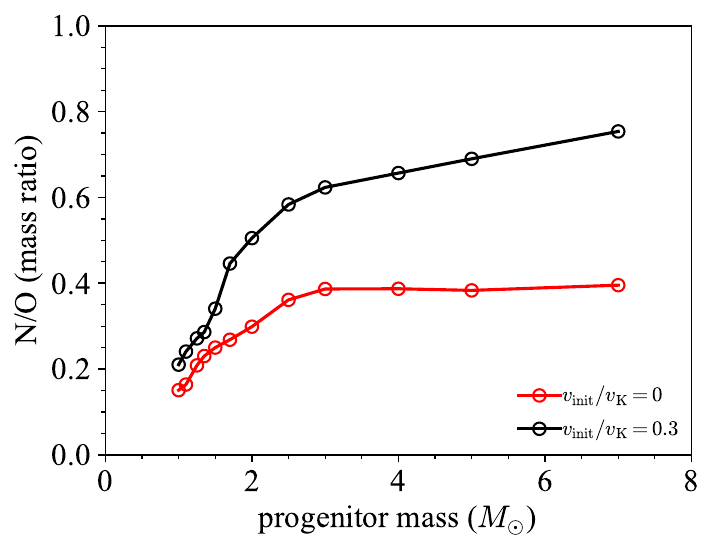}
 \end{center}
 \caption{Same as Figure~\ref{fig:abun_low},  but the x-axis is mass ratios.
 }\label{fig:mass_low}
\end{figure}

\begin{figure}[h!]
 \begin{center}
  \includegraphics[width=75mm]{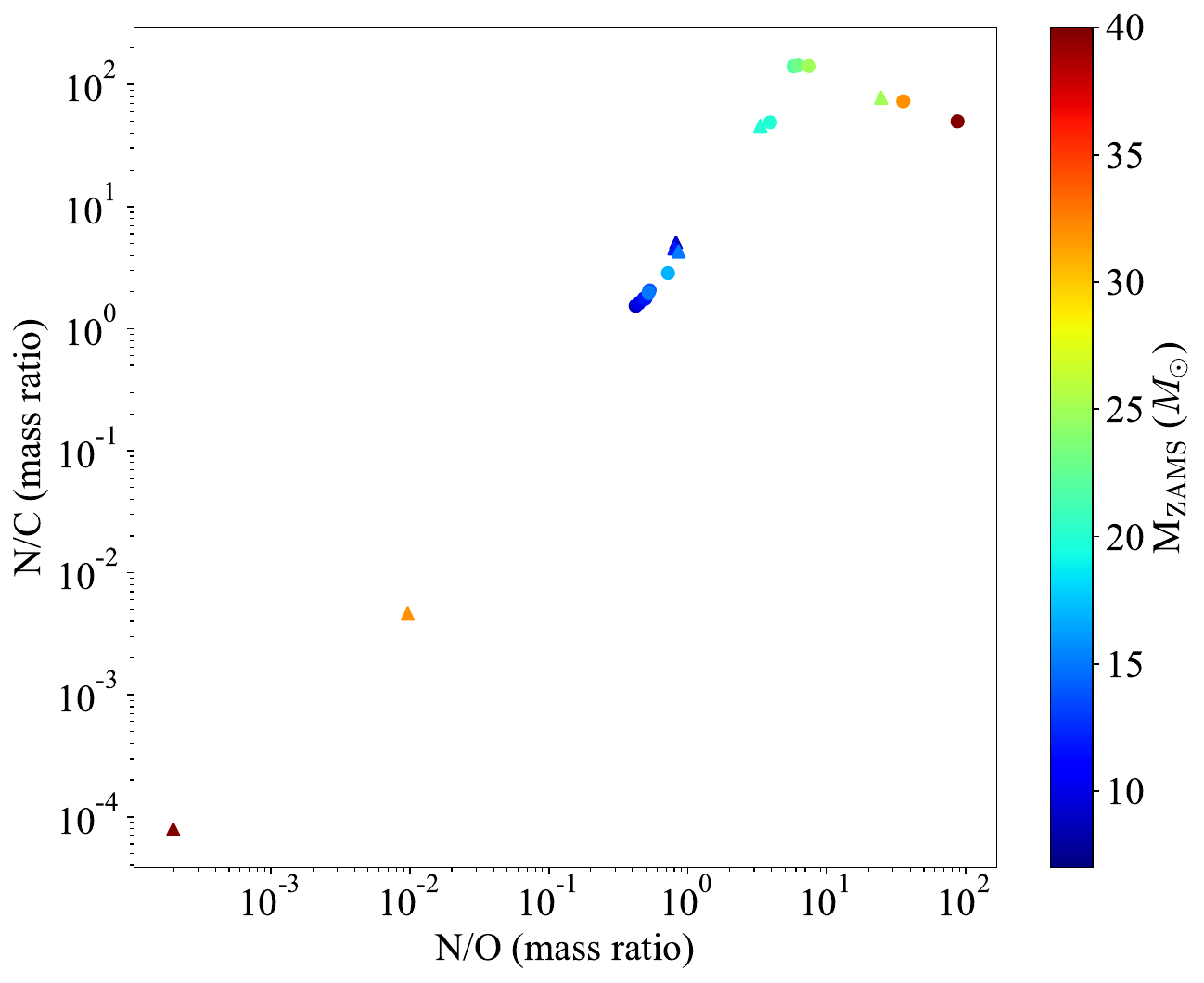}
  \includegraphics[width=75mm]{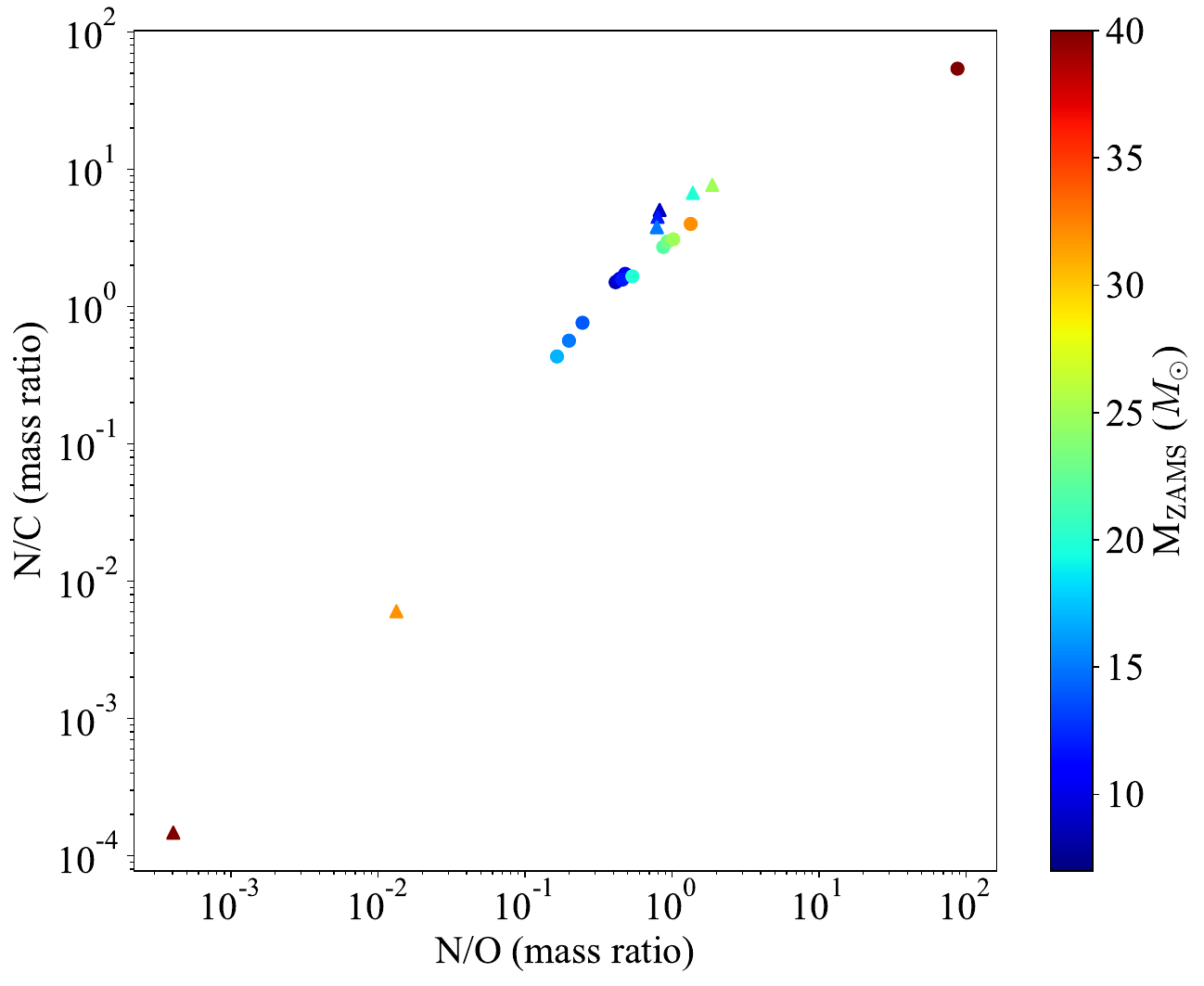}
  \includegraphics[width=75mm]{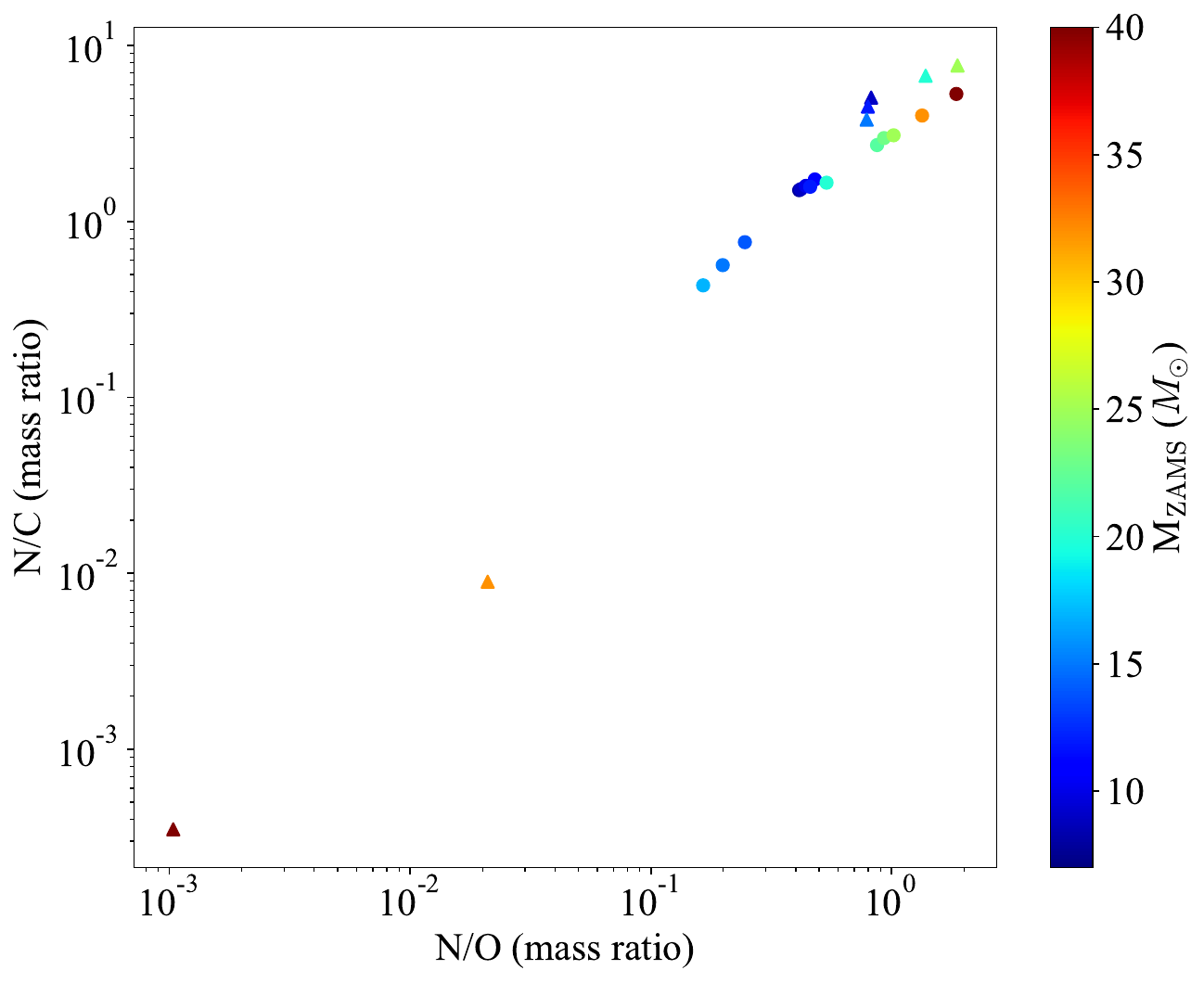}
 \end{center}
 \caption{Same as Figure~\ref{fig:ncno},  but for mass ratios.
 }\label{fig:ncno_mass}
\end{figure}

We here display the same plot as Figures~\ref{fig:abun}, \ref{fig:abun_low}, and \ref{fig:ncno} in which  N/C and N/O  are converted to those of the mass ratios (Figures~\ref{fig:mass}, \ref{fig:mass_low}, and \ref{fig:ncno_mass}).

\acknowledgments
The authors are very grateful to Takashi Yoshida, Jacco Vink, Satoru Katsuda, Hideyuki Umeda, Toshiki Sato, Kai Matsunaga, and Takaaki Tanaka for giving us precious advice on this study.
This work is supported by JSPS Core-to-Core Program (grant number: JPJSCCA20220002) and JSPS/MEXT Science Research grant Nos. JP19K03915, JP22H01265 (H.U.), and JP23KJ1350 (T.N.).

\bibliography{bibtex}{}
\bibliographystyle{JHEP}

%
%

\end{document}